\documentclass[12pt,preprint]{aastex}

\slugcomment{Accepted for publication in the Astrophysical Journal Letters}

\shorttitle{Self-consistent nonlinear force-free solution for an active 
region}
\shortauthors{Wheatland and R\'{e}gnier}

\begin{document}

\title{A self-consistent nonlinear force-free solution for a solar 
active region magnetic field}

\author{M.S. Wheatland}
\affil{Sydney Institute for Astronomy, School of Physics, 
  University of Sydney, NSW 2006, Australia}
\email{m.wheatland@physics.usyd.edu.au}

\and

\author{S. R\'{e}gnier}
\affil{School of Mathematics and Statistics, University of St Andrews,
St Andrews, Fife KY16 9SS, United Kingdom}
\email{stephane@mcs.st-andrews.ac.uk}

\begin{abstract}
Nonlinear force-free solutions for the magnetic field in the solar 
corona constructed using photospheric vector magnetic field boundary 
data suffer from a basic problem: the observed boundary data are 
inconsistent with the nonlinear force-free model. 
Specifically, there are two possible choices of boundary 
conditions on vertical current provided by the data, and the two 
choices lead to different force-free solutions. A novel solution 
to this problem is described. Bayesian probability is used 
to modify the boundary values on current density, using field-line
connectivity information from the two force-free solutions and taking 
into account uncertainties, so that the boundary data are more 
consistent with the two nonlinear force-free solutions. This procedure may 
be iterated until a set of self-consistent boundary data (the solutions 
for the two choices of boundary conditions are the same) is achieved.  
The approach is demonstrated to work in application to Hinode/SOT 
observations of NOAA active region 10953.
\end{abstract}


\section{Introduction}

Solar coronal magnetic fields provide the source of energy for solar
flares, and there is strong interest in developing methods 
for accurately modelling these fields, as a basis for improved 
understanding of solar activity. Spectro-polarimetric measurements 
of magnetically sensitive photospheric lines may be used to infer 
the vector magnetic field at the photosphere. In principle these 
measurements provide boundary values for modelling the overlying solar 
corona using magnetic field extrapolation techniques. However, in 
practice basic difficulties prevent the construction of reliable 
models (Schrijver et al.\ 2008; DeRosa et al.\ 2009).

A popular model for the coronal magnetic field in the low-density 
corona is the force-free model, involving a static balance of magnetic 
forces (e.g.\ Sakurai 1981; McClymont et al.\ 1997). The model is 
justified by the low ratio of gas pressure to magnetic pressure, or 
plasma beta, in the solar corona (e.g.\ Gary 2001). A nonlinear 
force-free magnetic field ${\bf B}$ satisfies 
$\nabla \times {\bf B} =\alpha {\bf B}$ 
and ${\bf B}\cdot \nabla \alpha =0$, 
where $\alpha$ is the force-free 
parameter, which is constant along field lines, but varies in 
space from field line to field line. The boundary 
conditions for the nonlinear problem consist of a specification of 
the normal component of ${\bf B}$ in the boundary (denoted $B_n$), 
together with a specification of $\alpha $ over one polarity (sign) of 
$B_n$~(e.g.\ Grad \& Rubin 1958; Sakurai 1981; Aly 1989; Amari et al.\ 
2006). The $\alpha$-boundary condition is equivalent to a 
specification of the normal component of the electric current density 
${\bf J}=\alpha {\bf B}/\mu_0$ over one polarity of $B_n$. 

Nonlinear force-free boundary value problems are difficult to solve 
in general. A number of numerical methods have been developed (for a
recent review, see Wiegelmann 2008), and demonstrated to work on test 
cases (e.g.\ Schrijver et al.\ 2006; Metcalf et al.\ 2008). Not all 
methods use the boundary conditions on $\alpha$ outlined above.  
For example, the optimization method 
(Wheatland et al.\ 2000; Wiegelmann 2004) and some versions of the 
magnetofrictional method (e.g.\ Roumeliotis 1996; Valori et al.\ 
2005) specify all three components of the vector magnetic field in the 
boundary over both polarities. Although in general this is an 
over-prescription, if the boundary values are consistent with the 
force-free model, this does not introduce a problem. A class of methods 
based on Grad \& Rubin (1958) use the boundary conditions on $\alpha$ 
described above, and the code employed in this paper is a Grad-Rubin, or 
``current-field iteration'' method (for details, see Wheatland 2007).

In two recent workshops (Schrijver et al.\ 2008; DeRosa et al.\ 
2009), a number of nonlinear force-free methods were critically 
assessed in application to solar vector magnetic field data from 
the Spectro-Polarimeter (SP) instrument of the Solar Optical Telescope 
(SOT) on the Hinode satellite (Tsuneta et al.\ 2008). Different 
methods were found to produce significantly different 
coronal field solution for the same active region, and in particular 
the magnetic energy of the different solutions varied substantially,
preventing reliable determination of the magnetic free energy of the
active region. The results from individual methods also lacked 
self-consistency. For example, for the Grad-Rubin methods
(Amari et al.\ 1997, 2006; Wheatland 2007), the two choices 
of polarity for the boundary conditions on electric current density 
led to different force-free solutions. Despite this basic problem, 
nonlinear force-free modelling is often applied to solar boundary data 
for selected active regions (e.g.\ R\'{e}gnier \& Priest 2007; Wang et 
al.\ 2008; Canou et al.\ 2009).

Determinations of the photospheric magnetic field transverse to the 
line of sight are substantially uncertain, and the errors are 
likely to contribute to the inconsistency problem. However, a more 
fundamental difficulty is that the magnetic field is unlikely to be 
force-free at the level of the measurements. The denser photospheric
plasma is subject to magnetic, pressure, gravity, and dynamical forces 
(Metcalf et al.\ 1995). Necessary conditions for a force-free 
field may be checked by calculating integrals of the field in the 
boundary representing the net magnetic flux, and the net force and 
torque on the field (Molodenskii 1969; Aly 1984, 1989). The integrals 
are zero for boundary data from a force-free field, but (in general) 
are found to be non-zero for solar photospheric data. One approach 
to the problem involves ``preprocessing'' the data to minimize these 
integrals (Wiegelmann et al.\ 2006). However, the conditions are 
necessary but not sufficient, and after preprocessing the boundary 
data are still inconsistent with the force-free equations (DeRosa et 
al.\ 2009).  Also, preprocessing typically involves smoothing the 
data, which is undesirable. 

An alternative approach to the problem is to calculate a force-free 
solution (or solutions) with boundary conditions which depart from 
the observed boundary data, and to then adjust the boundary conditions 
on the solution(s) until a ``best fit'' is achieved with the observed 
boundary data (e.g.\ Roumeliotis 1996; Aly \& Amari 2007). In this 
paper we demonstrate such a scheme. The method uses the information 
on field line connectivity provided by the two force-free solutions 
constructed from the two choices of boundary conditions on $\alpha$, 
and takes into account uncertainties in the $\alpha$ values. Bayesian 
probability (e.g.\ Jaynes 2003) is used to adjust the boundary values 
iteratively until a self-consistent set of values (the two force-free 
solutions are the same) is achieved.

The layout of the paper is as follows. Section~2 describes the 
method, and Section~3 presents a simple application to Hinode/SOT 
data for NOAA active region 10953, the subject of a recent nonlinear 
force-free workshop (DeRosa et al.\ 2009). Section~4 discusses the 
results. 

\section{Method}

The available solar data are assumed to be a set of values 
$(B_x,B_y,B_z)$ of the magnetic field over an observed region on
the photosphere. We neglect solar curvature, and assume $z$ is 
the vertical direction, and $z=0$ is the photospheric plane. 
Boundary values of the force-free parameter at $z=0$ may be 
obtained using $\alpha_0 =\mu_0J_z/B_z$, where
\begin{equation}\label{eq:jz}
\mu_0J_{z}=\frac{\partial B_y}{\partial x}
  -\frac{\partial B_x}{\partial y}
\end{equation}
is estimated by finite differences. Uncertainties
in the magnetic field components may be used to calculate 
corresponding uncertainties $\sigma_0$ in the estimates of
$\alpha_0$ (e.g.\ Leka \& Skumanich 1999). 

As described in 
Section~1, the values $\alpha_0$ together with the vertical 
component $B_z$ of the field provide two sets of boundary values for 
the force-free problem: one with $\alpha_0$ chosen on the 
positive polarity (denoted $P$), and one with $\alpha_0$ 
chosen on the negative polarity ($N$).
The current-field iteration method may be applied using the 
$\alpha_0$ values on polarity $P$, to give one nonlinear force-free
solution. Values of the force-free parameter are constant along field 
lines in a force-free model, so the solution maps values of 
$\alpha_0$ at points in $P$ to points in $N$, at the conjugate foot 
points of field lines. These mappings define new values $\alpha_1$ 
of the force-free parameter over points in $N$. 
The current-field iteration procedure may also be applied using 
the $\alpha_0$ values 
on the polarity $N$ as boundary conditions, to give a second 
nonlinear force-free solution. This solution maps the $\alpha_0$ 
values in $N$ to points in $P$. Hence it defines new values 
$\alpha_1$ of the force-free parameter at points in $P$. The result
of the two solutions is a complete set of values of $\alpha_1$
(i.e.\ defined over both $P$ and $N$). The new values also have 
associated uncertainties $\sigma_1$ obtained by mapping the uncertainty
values from the source polarity in each case. Hence at each point over
the observed region we have two possible sets of values: 
$(\alpha_0,\sigma_0)$ or  $(\alpha_1,\sigma_1)$.  

Bayes's theorem may be used to decide on a most probable 
single value of the force-free parameter at each boundary point. 
The theorem 
may be stated as 
${\mathcal P}(M|D,I)\propto {\mathcal P}(D|M,I){\mathcal P}(M,I)$, 
where ${\mathcal P}$ denotes a
probability, $M$ a model, $D$ data, and $I$ other information. In
our context $M$ is the value of $\alpha$ to be decided on,
$D$ is the new information from the mappings, i.e.\
($\alpha_1$, $\sigma_1$), and $I$ is the information available
before the mappings, i.e.\ ($\alpha_0$, $\sigma_0$).  Assuming Gaussian
errors we have
${\mathcal P}(D|M,I)\propto \exp [-(\alpha - \alpha_1)^2/(2\sigma_1^2)]$
and ${\mathcal P}(M,I)\propto \exp [-(\alpha - \alpha_0)^2/(2\sigma_0^2)]$.
Writing ${\mathcal L}(\alpha )=-\ln {\mathcal P}(M|D,I)$, we have
\begin{equation}
{\mathcal L}(\alpha )=\frac{(\alpha-\alpha_0)^2}{2\sigma_0^2}
+\frac{(\alpha-\alpha_1)^2}{2\sigma_1^2},
\end{equation}
ignoring an additive constant.
The most probable value of $\alpha$, which we denote by $\alpha_2$,
is then given by ${\mathcal L}^{\prime}(\alpha_2 )=0$:
\begin{equation}\label{eq:bayes_avg}
\alpha_2 = \frac{
  \alpha_0/\sigma_0^2+\alpha_1/\sigma_1^2}
              {1/\sigma_0^2+1/\sigma_1^2},
\end{equation}
i.e.\ an uncertainty-weighted average value. A corresponding 
uncertainty $\sigma_2$ may be defined assuming Gaussian behavior in the
vicinity of the peak by $\sigma_2=\left[{\cal
L}^{\prime\prime}(\alpha_2)\right]^{-1/2}$, yielding
\begin{equation}\label{eq:bayes_unc}
\sigma_2=\left(\frac{1}{\sigma_0^2}+\frac{1}{\sigma_1^2}\right)^{-1/2}.
\end{equation}
If the uncertainties in $\alpha_0$ (and hence 
also $\alpha_1$) are assumed to be equal at all points, then 
$\alpha_2=\frac{1}{2}\left(\alpha_0+\alpha_1\right)$, 
the simple average. 

The resulting values of $\alpha_2$ will still be inconsistent
with a force-free field, in general, but they are expected to be
closer to consistency. The process may then be repeated, using
the $\alpha_2$ values in the place of $\alpha_0$. Two solutions
are calculated, one from each polarity, and then the field line mappings 
of the solutions and the values $(\alpha_2,\sigma_2)$ define a new 
set of values $(\alpha_3,\sigma_3)$. Equations~(\ref{eq:bayes_avg})
and~(\ref{eq:bayes_unc}) are applied again, leading to a new set of 
values $(\alpha_4,\sigma_4)$.  This process is iterated. The construction 
of a pair of solutions, and their use to obtain a new $\alpha$-map, 
represents a `self-consistency cycle.' For convenience we label the 
two force-free solutions constructed during each cycle by an index $k$, so 
that the first cycle involves solution numbers $k=1$ and $k=2$. 
It is expected that the 
procedure will converge after a number of self-consistency cycles, in 
the sense that the two solutions from the different polarities become 
identical. The result is expected to be a single force-free 
solution, with a minimum departure from the observations. 

\section{Application to Hinode/SOT data}

To demonstrate the method, we consider Hinode/SOT data from the
recent force-free workshop, for NOAA active region 10953, observed at 
22:30~UT on 30 April 2007. The data are described
in  DeRosa et al.\ (2009), and consist of field components 
$(B_x,B_y,B_z)$ on a $320\times 320$ grid spanning a $185.6\,{\rm Mm}$
square area. The Hinode data fill only part of the $320\times 320$ 
field of view, with the additional $B_z$ values derived from a 
line-of-sight magnetogram from the Michelson Doppler Interferometer 
(MDI) instrument on the Solar and Heliospheric Observatory (SOHO) 
spacecraft (Scherrer et al.\ 1995). The data used here are not 
preprocessed, and no smoothing is applied. Values of $\alpha_0$ 
are obtained by centred differencing of $B_x$
and $B_y$ values according to Equation~(\ref{eq:jz}), for all points
with $|B_z|>0.01\times \mbox{max}(B_z)$. Values of $\alpha_0$ 
are zero for points in the field of view corresponding to the MDI
data. Uncertainties are not available for the photospheric field
measurements, so we assume that the uncertainties in the $\alpha_0$
values are equal at all points.

The current-field iteration method (Wheatland 2007) is used to 
calculate force-free solutions from the boundary values for $\alpha$ 
on the two polarities, for 10 self-consistency cycles. Each solution 
involves 20 Grad-Rubin iterations, sufficient to achieve approximate
convergence. The solutions are constructed on a $320\times 320\times 256$
grid. During the construction of each solution, points on the
grid threaded by field lines which cross the sides or top boundaries
of the computational volume (including points on the grid in the lower 
boundary) have $\alpha$ set to zero. This provides a simple solution 
to the problem of ``missing information'' associated with the absence 
of boundary conditions on the sides and top of the computational 
volume. After each cycle, a new $\alpha$-map is constructed according 
to Equation~(\ref{eq:bayes_avg}), with the assumption of constant
uncertainties.

The procedure is found to converge: the fields constructed from the two 
choices of boundary conditions are very similar after 10 cycles. Figure~1 
illustrates field lines for the two solutions at the first and the last 
cycles. The left panel shows the two force-free fields at the first cycle, 
constructed from the original boundary 
data. The blue solution ($k=1$) uses $\alpha_0$ values on the positive
polarity, and the red solution ($k=2$) uses $\alpha_0$ values on the
negative polarity [this is similar to the ``Wh$^{-}$'' solution from 
DeRosa et al.\ (2009), except that the Wh$^{-}$ solution used preprocessed
boundary data]. The blue and red field lines are quite different. The
red field lines are more distorted, suggesting that this solution
is more non-potential. The right 
panel shows the corresponding solutions at the tenth cycle (solutions 
$k=19$ and $k=20$). The two solutions are very similar. The image in the 
background of each panel shows the boundary values of $B_z$, with positive
polarity areas appearing light, and negative polarity dark.
 
Figure~2 illustrates two quantitative measures of the difference between
the two solutions at each cycle. The upper panel shows the magnetic energy 
$E_k$ of each solution, in units of the energy $E_0$ of the potential field. 
The field constructed from the values of $\alpha_0$ on the negative
polarity ($k=2$) has substantially more energy than the field constructed
using the positive polarity ($k=1$), as suggested by the appearance of 
the field lines in Figure~1. After 10 self-consistency cycles the energies 
of the two solutions are very similar (they differ by $<0.03\%$). The 
dimensional energy of the potential field is 
$E_0=8.96\times 10^{25}\,{\rm J}$, and the energy of the final fields 
obtained by the self-consistency procedure is
$E_f=9.11\times 10^{25}\,{\rm J}$. Hence the free energy of the magnetic
field for the final solutions is $E=1.5\times 10^{24}\,{\rm J}$.
The lower panel in Figure~2 shows the mean vector error between 
solutions $k$ and $k-1$, defined by (Schrijver et al.\ 2006):
\begin{equation}\label{eq:L2}
\mbox{MVE}_k
  =\frac{1}{N_xN_yN_z}\sum_i
  \frac{|{\bf B}^{(k)}_i-{\bf B}^{(k-1)}_i|}
  {|{\bf B}^{(k-1)}_i|},
\end{equation}
where $i$ runs over the points on the
computational grid. The mean vector error is reduced by more than a factor 
of 60 by the procedure.

It is also interesting to examine the changes in the boundary conditions
on current, and in the boundary components $B_x$ and $B_y$. Figure~3
shows the vertical electric current density $J_z=\alpha B_z/\mu_0$ at the
first and last cycles. The left panel shows the observed values
of $J_z$, and the right panel shows the values after the last cycle.
The currents have been reduced in magnitude overall
by the averaging in the self-consistency procedure, but it is 
notable that basic structures present in the original data remain. The 
changes in the horizontal field are substantial: the RMS change 
in $B_x$ across the entire field of view is $120\,{\rm G}$, and the RMS
change in $B_y$ is $100\,{\rm G}$. This is to be expected given the 
gross discrepancy between the two initial solutions (left panel in 
Figure~1). Some part of the change is due to the artificial construction 
of the boundary data, specifically the embedding of the Hinode/SOT data 
within a set of SOHO/MDI data with a larger field of view. Boundary points 
corresponding to the MDI data have $\alpha_0=0$, but during the 
self-consistency procedure non-zero values of $\alpha$ may be mapped to 
these points, leading to significant changes in $B_x$ and $B_y$.
For comparison, we note that the preprocessing procedure used at the recent
nonlinear force-free workshop (DeRosa et al.\ 2009) introduced RMS changes 
in $B_x$ and $B_y$ of about $60\,{\rm G}$, as well as an RMS change in 
$B_z$ of about $80\,{\rm G}$ (values of $B_z$ are unchanged in the
self-consistency procedure).

\section{Discussion}

A method for calculating a self-consistent nonlinear
force-free solution from solar photospheric vector magnetic field boundary
data is described. The ``self-consistency'' procedure resolves 
the fundamental problem that the solar data define two different 
force-free solutions. The method involves constructing the two solutions 
and then adjusting the boundary conditions on the force-free parameter
using the field line connectivity defined by the solutions, taking into 
account observational uncertainties. Iteration of the procedure leads 
to a set of boundary data for which there is only one force-free 
solution. The method is demonstrated to work in application to Hinode/SOT 
data for NOAA active region 10953.

The results for active region 10953 should be regarded as providing
a proof of concept, rather than as the construction of a 
completely realistic model, due to a number of limitations. For example, 
the results could be improved by assigning uncertainties to the $\alpha$ 
values, so that more reliable boundary conditions on the electric current 
density are treated preferentially. It is difficult to assign uncertainty
estimates which are correct in an absolute sense. An advantage of the
present method is that, according to Equation~(\ref{eq:bayes_avg}),
only the relative sizes of the uncertainties need to be correct. 
Another limitation of the present calculation is the embedding of 
Hinode/SOT data in a larger (SOHO/MDI) field of view for which the 
boundary values of $\alpha$ are zero. For field lines with one foot point
in the MDI region, the averaging in the self-consistency procedure 
reduces $|\alpha|$ by comparison the value at the Hinode/SOT foot 
point. It is likely that these limitations account in part for the 
relatively small free energy of the final field.

The self-consistency procedure provides a way to use the observed 
boundary information on electric currents from both polarities,
which is preferable to discarding the information from one polarity. 
It should be noted that the magnetofrictional method (e.g.\ Valori et 
al.\ 2005) and the optimization method (e.g.\ 
Wiegelmann 2004) already use the boundary data on currents 
from both polarities. However, the difference between the 
magnetofrictional/optimization approaches and the present method is
that the present method determines a force-free solution and a set 
of boundary conditions that are consistent with the force-free
model. If the observed boundary data are inconsistent with the
force-free model the magnetofrictional/optimization methods will not
achieve this, without additional modification [see Aly \& Amari (2007) 
for a suggested approach, in the context of the optimization method]. 

In future work the self-consistency procedure will be investigated
in more detail, via application to test cases with inconsistent boundary 
data. This should permit characterisation of the convergence of the 
scheme. Other solar (and solar-like) data will also be examined,
including the other cases from the nonlinear force-free workshops 
(Metcalf et al.\ 2008; Schrijver et al.\ 2008), and the method will be
applied with uncertainties when available.

The self-consistency procedure is applicable to a wealth of archival 
data from ground-based observatories, including the National Solar 
Observatory's Synoptic Optical Long-term Solar Investigations of the 
Sun Vector Spectromagnetograph (SOLIS/VSM), as well as satellite data 
from Hinode/SOT and the Helioseismic and Magnetic Imager (HMI) on NASA's 
Solar Dynamics Observatory (SDO), due to be launched in 2009. 
Reliable coronal magnetic field models should greatly enhance the
science value of data obtained from these facilities, permitting
investigation of diverse aspects of the physics of the solar 
atmosphere and of solar activity.

\acknowledgments

Stephane R\'{e}gnier acknowledges the support of a Royal Society 
Short Visit Grant during a visit to the University of Sydney.

{\it Facilities:} \facility{Hinode}, \facility{SOHO}.

\clearpage

\clearpage

\begin{figure}
\plotone{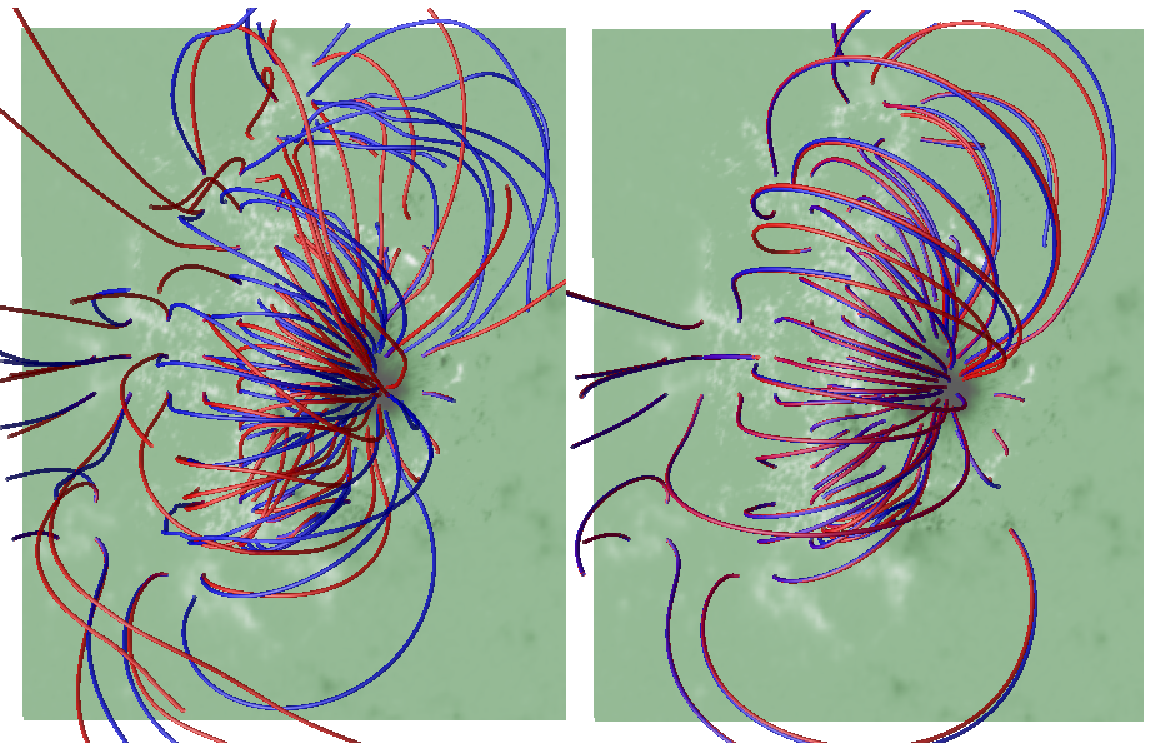}
\caption{Left panel: Overlay of field lines for the two force-free 
solutions constructed using boundary conditions on current density taken 
from the positive polarity (blue lines), and from the negative 
polarity (red lines), using the original boundary data (solutions $k=1$ and
$k=2$). Right panel: 
Overlay of the field lines for the two solutions after 10 self-consistency 
cycles (solutions $k=19$ and $k=20$). The image in the background of each
panel shows the boundary values of $B_z$, with positive polarity areas
appearing light, and negative polarity areas dark. See the electronic 
edition of the Journal for a color version 
of this figure.\label{fig1}}
\end{figure}

\begin{figure}
\epsscale{0.80}
\plotone{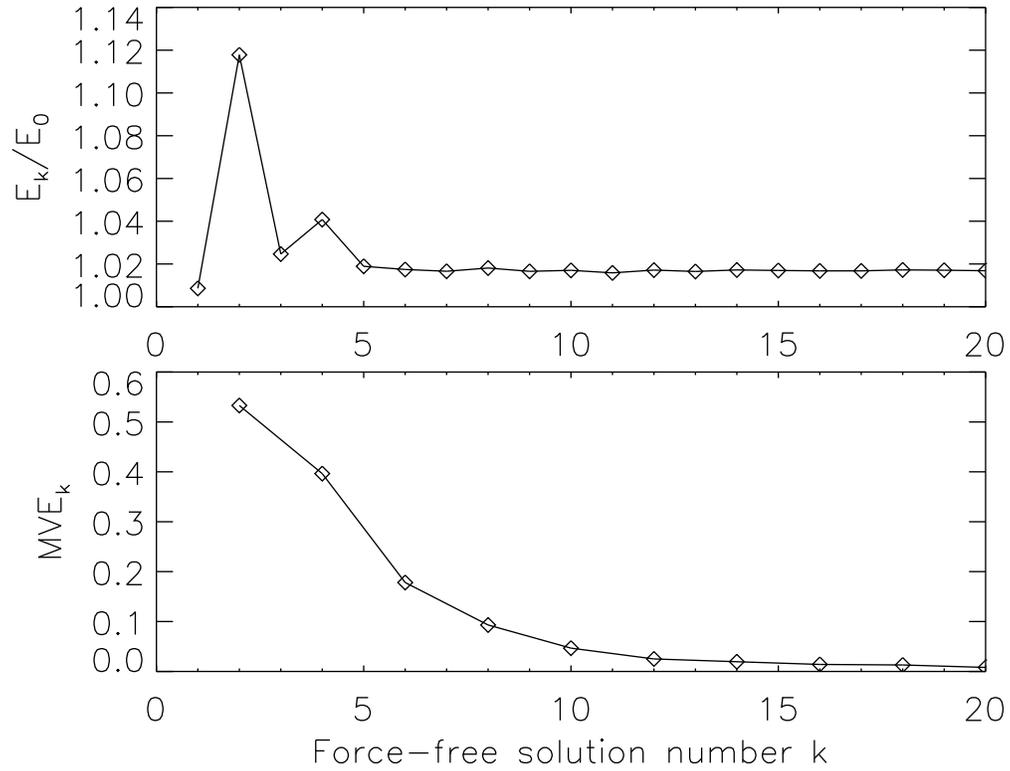}
\caption{Upper panel: The magnetic energy $E_k$ of each force-free 
solution, in units of the energy of the potential field $E_0$, over 
the 10 self-consistency
cycles. Solutions $k=1,3,...,19$ are constructed using $\alpha$ values on the
positive polarity, and solutions $k=2,4,...,20$ using values on the negative
polarity. Lower panel: The mean vector error, quantifying the 
discrepancy between the two fields constructed at each cycle.}. 
\end{figure}

\begin{figure}
\epsscale{1.0}
\plotone{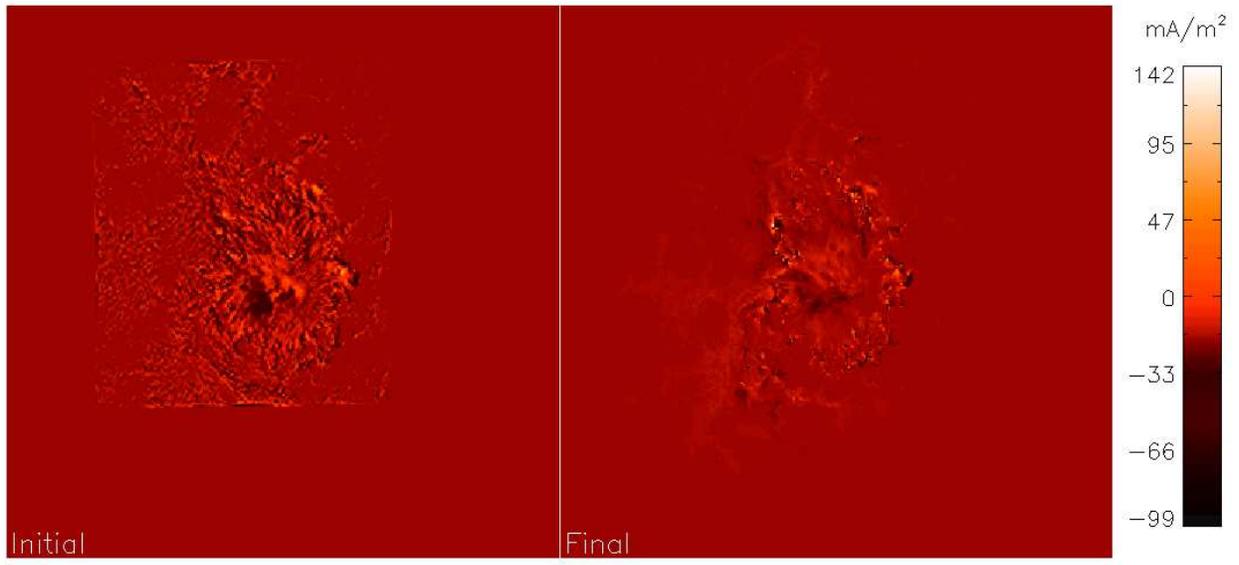}
\caption{Boundary conditions on electric current density in the 
observations (left panel) and in the solution after 10 self-consistency 
cycles (right panel).  
}
\end{figure}

\end{document}